\documentclass[%
 reprint,
 amsmath,amssymb,
 aip,
apl
]{revtex4-1}

\usepackage{graphicx}
\usepackage{dcolumn}
\usepackage{bm}
\usepackage{upgreek} 

\begin{document}


\title{Exchange magnetic field torques in YIG/Pt bilayers observed by the spin-Hall magnetoresistance}

\author{N. Vlietstra}
\affiliation{ 
Physics of Nanodevices, Zernike Institute for Advanced Materials, University of Groningen, Groningen, The Netherlands
}%

\author{J. Shan}
\affiliation{ 
Physics of Nanodevices, Zernike Institute for Advanced Materials, University of Groningen, Groningen, The Netherlands
}%

\author{V. Castel}
\affiliation{ 
Physics of Nanodevices, Zernike Institute for Advanced Materials, University of Groningen, Groningen, The Netherlands
}%

\author{J. Ben Youssef}%
\affiliation{ 
Laboratoire de Magn\'etisme de Bretagne, CNRS, Universit\'e de Bretagne Occidentale, Brest, France
}%

\author{G. E. W. Bauer}
\affiliation{
Kavli Institute of NanoScience, Delft University of Technology, Delft, The Netherlands }
\affiliation{Institute for Materials Research and WPI-AIMR, Tohoku University, Sendai, Japan 
}

\author{B. J. van Wees}
\affiliation{ 
Physics of Nanodevices, Zernike Institute for Advanced Materials, University of Groningen, Groningen, The Netherlands
}%

\date{\today}

\begin{abstract}

The effective field torque of an yttrium-iron-garnet film on the spin accumulation in an attached Pt film is measured by the spin-Hall magnetoresistance (SMR). As a result, the magnetization direction of a ferromagnetic insulating layer can be measured electrically. Experimental transverse and longitudinal resistances are well described by the theoretical model of SMR in terms of the direct and inverse spin-Hall effect, for different Pt thicknesses [3, 4, 8 and 35nm]. Adopting a spin-Hall angle of Pt $\theta_{SH}=0.08$, we obtain the spin diffusion length of Pt ($\lambda=1.1\pm0.3$nm) as well as the real ($G_r=(7\pm3)\times10^{14}\Omega^{-1}$m$^{-2}$) and imaginary part ($G_i=(5\pm3)\times10^{13}\Omega^{-1}$m$^{-2}$) of the spin-mixing conductance and their ratio ($G_r/G_i=16\pm4$). 

\end{abstract}

\keywords{spin-Hall magnetoresistance, yttrium iron garnet, YIG, spin-mixing conductance, effective field torque}

\maketitle


In spintronics, interfaces between magnets and normal metals are important for the creation and detection of spin currents, which is governed by the difference of the electric conductance for spin up and spin down electrons.\citep{HeinrichInterface,TserkovnyakPRB2002,SaitohIEEE} Another important interaction between the electron spins in the magnetic layer and those in the normal metal, that are polarized perpendicular to the magnetization direction, is governed by the spin-mixing conductance $G_{\uparrow\downarrow}$,\citep{BauerMix} which is composed of a real part and an imaginary part ($G_{\uparrow\downarrow}=G_r+i G_i$). 
$G_r$ is associated with the ``in-plane'' or ``Slonczewski'' torque along $\vec{m}\times\vec{\mu}\times\vec{m}$,\citep{STTintro,WangAPL2011,Kajiwara2010nature} where $\vec{m}$ is the direction of the magnetization of the ferromagnetic layer and $\vec{\mu}$ is the polarization of the spin accumulation at the interface. $G_i$ describes an exchange magnetic field that causes precession of the spin accumulation around $\vec{m}$. This ``effective-field'' torque associated with $G_i$ points towards $\vec{\mu}\times\vec{m}$. 

While several experiments succeeded in measuring $G_r$,\citep{BauerMix,GerritMixCon,Kajiwara2010nature,CzeschkaScalingBehavior,SaitohIEEE,CastelThicknessPt} $G_i$ is difficult to determine experimentally, mainly because it is usually an order of magnitude smaller than $G_r$.\citep{BauerMix} The recently discovered spin-Hall magnetoresistance (SMR)\citep{BauerTheorySMR,VlietstraSMR,BauerSMR,JamalTaPt} offers the unique possibility to measure $G_i$ for an interface of a normal metal and a magnetic insulator by exposing it to out-of-plane magnetic fields. Althammer \emph{et al.}\citep{AlthammerSMR} carried out a quantitative study of the SMR of Yttrium Iron Garnet (YIG)/Platinum (Pt) bilayers. They obtained an estimate of $G_i=1.1\times10^{13}\Omega^{-1}$m$^{-2}$ by extrapolating the high field Hall resistances to zero magnetic field.\footnote{The authors of Ref. \citep{AlthammerSMR} obtained $G_i$ by adding the saturation magnetization to the applied magnetic field to obtain the total magnetic field in the Pt. In our opinion the saturation magnetization should not be included, which leads to a different zero-field extrapolation resulting in $G_i=1.7\times10^{13}\Omega^{-1}$m$^{-2}$, which is more close to the uncertainty interval of our results.}

In this paper, we report experiments in which the contribution of $G_r$ and $G_i$ can be controlled the magnetization direction of the YIG layer by an external magnetic field. Thereby either $G_r$ or $G_i$ can be made to dominate the SMR. By fitting the experimental data by the theoretical model for the SMR,\citep{BauerTheorySMR} the magnitude of $G_r$, $G_i$ and the spin diffusion length $\lambda$ in Pt are determined. 
For SMR measurements, Pt Hall bars with thicknesses of 3, 4, 8 and 35nm were deposited on YIG by dc sputtering.\citep{VlietstraSMR} Simultaneously, a reference sample was fabricated on a Si/SiO$_2$ substrate. The length and width of the Hall bars are 800$\upmu$m and 100$\upmu$m, respectively. The YIG has a thickness of 200nm and is grown by liquid phase epitaxy on a single crystal Gd$_3$Ga$_4$O$_{12}$ (GGG) substrate.\citep{CastelPRB} The magnetization of the YIG has an easy-plane anisotropy, with an in-plane coercive field of only 0.06mT. To saturate the magnetization of the YIG in the out-of-plane direction a field above the saturation field $B_s$ ($\mu_0M_s=0.176$T)\citep{CastelPRB} has to be applied. All measurements are carried out at room temperature.

\begin{figure}
\includegraphics[width=8.5cm]{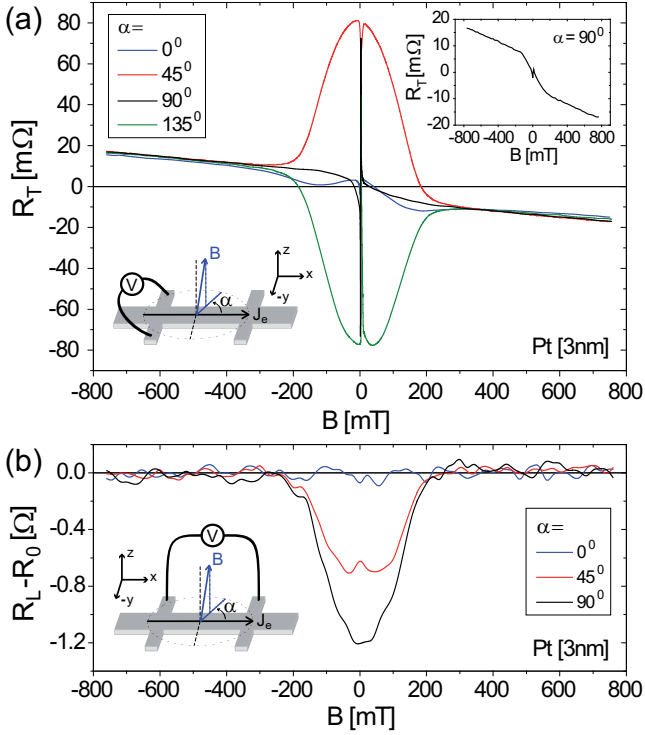}
\caption{\label{fig:Fig1} 
(a) Transverse and (b) longitudinal resistance of Pt [3nm] on YIG under an applied out-of-plane magnetic field. $\alpha$ is the angle between $J_e$ and the small in-plane component of the applied magnetic field. The insets show the configuration of the measurements, as well as a separate plot of the transverse resistance for $\alpha=90^{\circ}$, where the contribution of $G_i$ is most prominent. $R_0$ is the high-field resistance of the Pt film, here 1695$\Omega$.
}
\end{figure}

The magnetization of the YIG is controlled by sweeping the out-of-plane applied magnetic field with a small intended in-plane component (see insets of Fig. \ref{fig:Fig1}(a,b)). 
Fig. \ref{fig:Fig1}(a) shows out-of-plane magnetic field sweeps for various directions of the in-plane component of $B$ (and thus $M$), while measuring the transverse resistance (using a current $I=1$mA). Above the saturation field ($B>B_s$), a linear magnetic field dependence is observed, that can be partly ascribed to the ordinary Hall effect, but its slope is slightly larger, which suggests the presence of another effect (discussed below). Furthermore, extrapolation of the linear regime for the positive and negative saturated fields to $B=0$mT, reveals an offset between both regimes, that, as shown below, can be ascribed to $G_i$.
When $B$ is smaller than the saturation field, the observed signal strongly depends on the angle $\alpha$ between the direction of the charge current $J_e$ and the in-plane component of the magnetic field. This $\alpha$-dependence is not observed for $B>B_s$. The maximum/minimum magnitude of the peak/dip observed in the non-saturated regime exactly follows the SMR behaviour for in-plane magnetic fields.\citep{BauerSMR,VlietstraSMR} By increasing the magnetic field strength, the magnetization is tilted out of the plane and less charge current is generated by the inverse spin-Hall effect in the transverse (and also longitudinal) direction, resulting in a decrease of the SMR signal. The sharp peak observed around zero applied field can be explained by the reorientation of $M$ in the film plane when $B$ is swept through the coercive field of the YIG.

The corresponding measurements of the longitudinal resistance are shown in Fig. \ref{fig:Fig1}(b) (For currents $I=1-100\upmu$A). In this configuration, the signal for $B>B_s$ does not show a field dependence nor an offset between positive and negative field regimes when linearly extrapolated to zero field. 

The observed features for the transverse (Fig. \ref{fig:Fig1}(a)) as well as the longitudinal (Fig. \ref{fig:Fig1}(b)) resistance can be described by the following equations\citep{BauerTheorySMR}
\begin{equation} \label{eq:rhoT}
	\rho{_T}=\Delta\rho{_1}m_x m_y+\Delta\rho{_2}m_z+(\Delta\rho{_{Hall}}+\Delta\rho{_{add}})B_z \\
\end{equation}
\begin{equation} \label{eq:rhoL}
	\rho{_L}=\rho+\Delta\rho{_0}+\Delta\rho{_1}(1-m{^2_y})  
\end{equation}
where $\rho{_T}$ and $\rho{_L}$ are the transverse and longitudinal resistivity, respectively. $\rho$ is the electrical resistivity of the Pt. $\Delta\rho{_{Hall}}B_z$ describes the change in resistivity caused by the ordinary Hall effect and $\Delta\rho{_{add}}B_z$ is the additional resistivity change on top of $\Delta\rho{_{Hall}}B_z$, as observed for saturated magnetic fields.\footnote{From the measurements for $\alpha=90^{\circ}$, shown in the inset of Fig. \ref{fig:Fig1}(a) and in Figs. \ref{fig:Fig3}(a)-(c), we deduce that also in the non-saturated regime this additional effect likely scales linearly with B. The dominant linear effect observed in the non-saturated regime is attributed to $G_i$. The remaining linear signal is explained by the sum of the ordinary hall effect and the additional term as defined in Eq. (\ref{eq:rhoT}).} $B_z$ is the magnetic field in the $z$-direction. $m_x$, $m_y$ and $m_z$ are the components of the magnetization in the $x$-, $y$- and $z$-direction, respectively, defined by $m_x=\cos\alpha \cos\beta$, $m_y=\sin\alpha \cos\beta$ and $m_z=\sin\beta$, where $\alpha$ is the in-plane angle between the applied field $B$ and $J_e$, and $\beta$ is the angle by which $M$ is tilted out of the plane. For an applied field in the $z$-direction, from the Stoner-Wohlfarth Model,\citep{StonerModel} $\beta=\arcsin{B/B_s}$. $\Delta\rho{_0}$, $\Delta\rho{_1}$ and $\Delta\rho{_2}$ are resistivity changes as defined below\citep{BauerTheorySMR}

\begin{align}
	\frac{\Delta\rho_0}{\rho}&= -\theta{^2_{SH}}\frac{2\lambda}{d{_N}} \tanh \frac{d_N}{2\lambda} \\
		\label{eq:rho1} 
		\frac{\Delta\rho_1}{\rho}&= \theta^2_{SH} \frac{\lambda}{d{_N}} {\rm Re} \left(\frac{2\lambda G_{\uparrow\downarrow}\tanh^2 \frac{d_N}{2\lambda}}{\sigma+2\lambda G_{\uparrow\downarrow}\coth\frac{d_N}{\lambda}}\right)  \\
 	\label{eq:rho2} 
 	\frac{\Delta\rho_2}{\rho}&= -\theta^2_{SH} \frac{\lambda}{d{_N}} {\rm Im} \left(\frac{2\lambda G_{\uparrow\downarrow}\tanh^2 \frac{d_N}{2\lambda}}{\sigma+2\lambda G_{\uparrow\downarrow}\coth\frac{d_N}{\lambda}}\right) 
\end{align}
where $\theta_{SH}$, $\lambda$, $d_N$, $G_{\uparrow\downarrow}$ and $\sigma$ are the spin-Hall angle, the spin relaxation length, the Pt thickness, the spin-mixing conductance ($G_{\uparrow\downarrow}=G_r+i G_i$) and the bulk conductivity, respectively.

\begin{figure}[b]
\includegraphics[width=8.5cm]{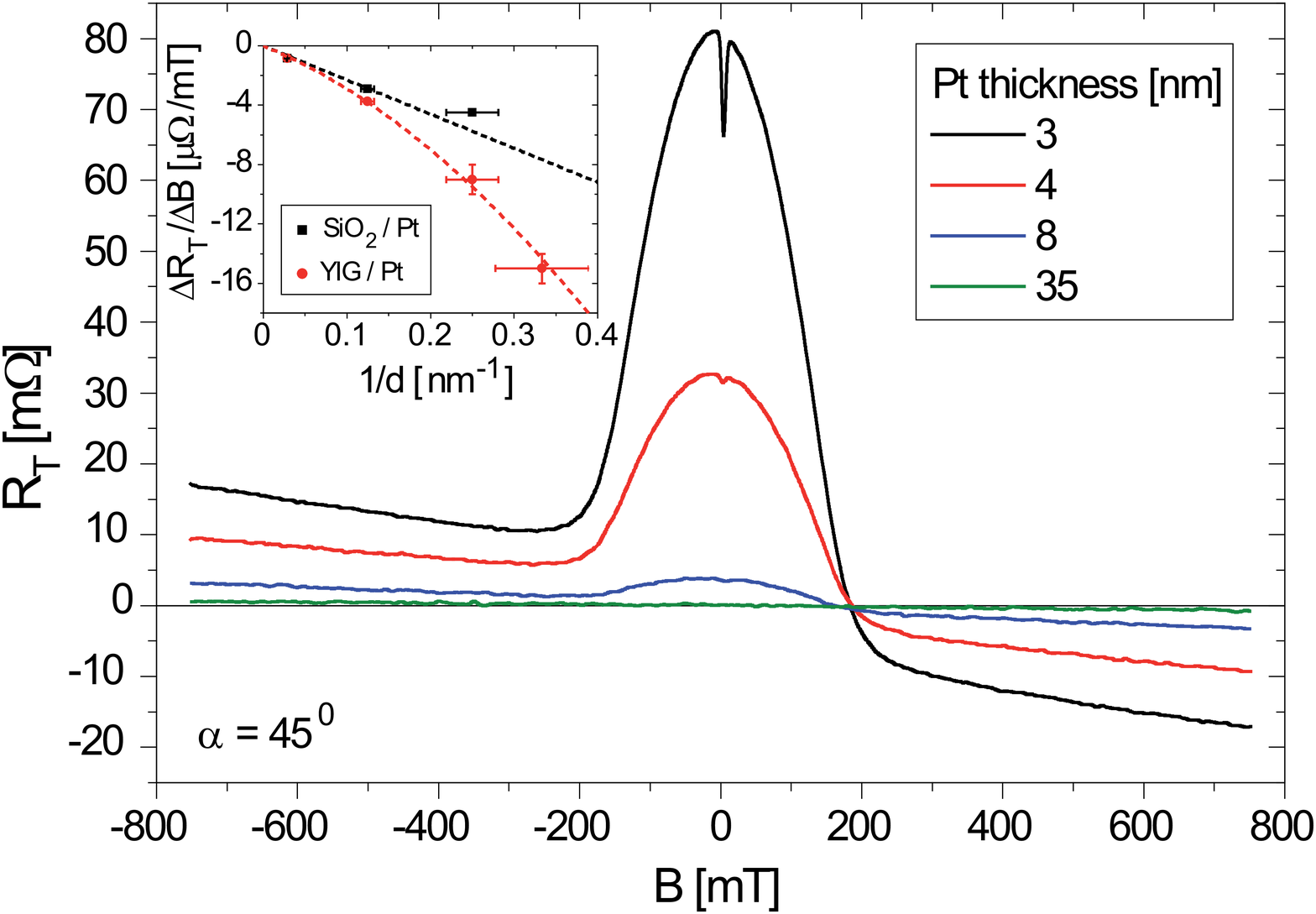}
\caption{\label{fig:Fig2} 
Out-of-plane magnetic field sweeps on YIG/Pt for different Pt thicknesses [3, 4, 8 and 35nm], fixing $\alpha=45^{\circ}$. In the saturated regime ($B>B_s$), linear behaviour is observed. The inset shows the measured slope $\Delta R_T / \Delta B$ in the saturated regimes (red dots). The expected (black line) and measured (black dots) curves display the slopes for the ordinary Hall effect on a SiO$_2$/Pt sample. The red dotted line is a guide for the eye.
}
\end{figure}

From Eq. (\ref{eq:rhoT}), $G_i$ is most dominant in the transverse configuration when the product $m_xm_y$ vanishes ($\Delta\rho{_2}$ is a function of $G_i$). This is the case for $\alpha=0^{\circ}$ and $\alpha=90^{\circ}$, as is shown in Fig. \ref{fig:Fig1}(a). As $m_z$ scales linearly with $B$, the term $\Delta\rho{_2}m_z$, contributes an additional linear dependence for $B<B_s$ that causes an offset between resistances for positive and negative saturation fields. This behaviour is clearly observed in the inset of Fig. \ref{fig:Fig1}(a), where the measurement for $\alpha=90^{\circ}$ is separately shown. For $\alpha=45^{\circ}$ (135$^{\circ}$), the product $m_xm_y$ is maximized (minimized) and a maximum (minimum) change in resistance is observed.

These measurements were repeated for a set of samples with different Pt thicknesses [3, 4, 8 and 35nm]. Results of the thickness dependent transverse resistance are shown in Fig. \ref{fig:Fig2}. For $\alpha=45^{\circ}$, at which both $G_r$ and $G_i$ contribute to a maximum SMR signal, a clear thickness dependence is observed at all field values. 
The thickness dependence of the slope $\Delta R_T / \Delta B$ at saturation fields is shown in the inset of Fig. \ref{fig:Fig2}, where the red dots represent the experiments. The black line (dots) shows the expected (observed) slope from the ordinary Hall effect (measured on a SiO$_2$/Pt sample) given by the equation $(\Delta R_T / \Delta B)_{Hall}=R_H/d_N$, where $R_H=-0.23\times10^{-10}$ m$^3$/C is the Hall coefficient of Pt.\citep{Hurd} $\Delta R_T / \Delta B$ for YIG/Pt behaves distinctively different. When decreasing the Pt thickness, $\Delta R_T / \Delta B$ of YIG/Pt increases faster than expected from the ordinary Hall effect. This discrepancy cannot be explained by the present theory for the SMR and may thus indicate a different proximity effect. The red dotted line in the inset of Fig. \ref{fig:Fig2} is a guide for the eye and represents the term $\Delta\rho{_{Hall}}+\Delta\rho{_{add}}$ in Eq. (\ref{eq:rhoT}). 

The SMR, including the resistance offset obtained by linear extrapolation of the high field regimes, is only significant for the thin Pt layers [3, 4 and 8nm]. The thick Pt layer [35nm] shows no (or very small) SMR.

\begin{figure}
\includegraphics[width=8.5 cm]{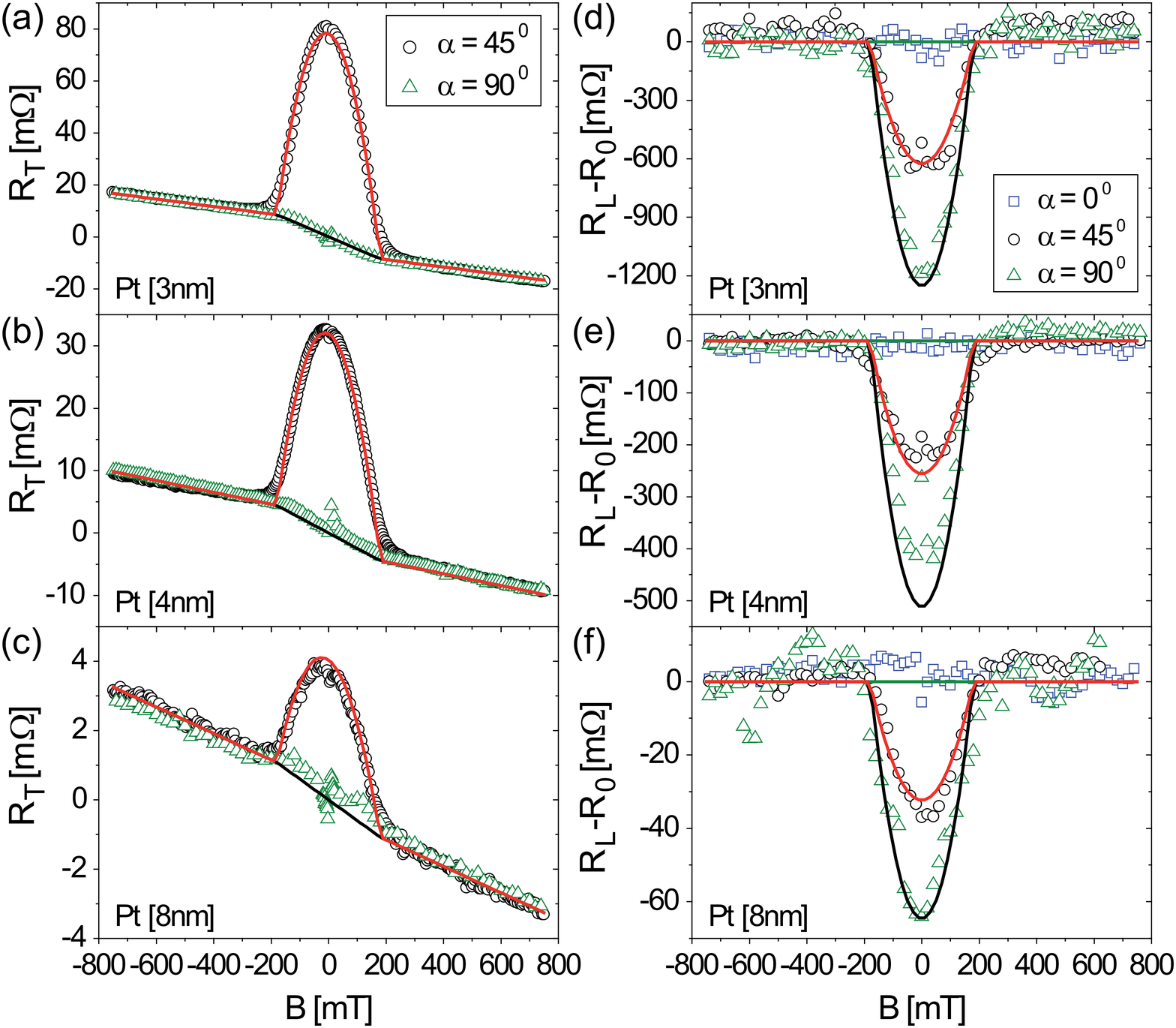}
\caption{\label{fig:Fig3} 
Theory Eqs. (\ref{eq:rhoT},\ref{eq:rhoL}) (solid lines) fitted to (a)-(c) transverse and (d)-(f) longitudinal observed resistances (open symbols) for different $\alpha$ and Pt thicknesses 3, 4 and 8nm, respectively, using $\theta_{SH}=0.08$, $\lambda=1.2$nm, $G_r=4.4\times10^{14}\Omega^{-1}$m$^{-2}$ and $G_i=2.8\times10^{13}\Omega^{-1}$m$^{-2}$. $R_0$ is the high-field longitudinal resistance of the Pt film of 1695$\Omega$, 930$\Omega$ and 290$\Omega$ for the 3, 4 and 8nm Pt thickness, respectively. 
}
\end{figure}

\begin{figure}
\includegraphics[width=8.5cm]{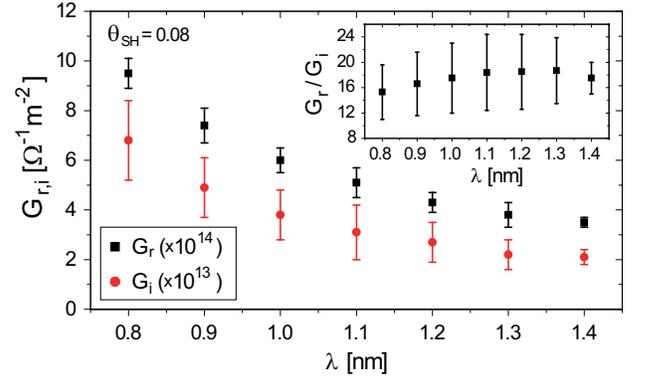}
\caption{\label{fig:Fig4} 
Obtained magnitude and uncertainties of $G_r$ and $G_i$ ($G_r/G_i$ in the inset) as a function of $\lambda$, for $\theta_{SH}=0.08$.
}
\end{figure}

Using Eqs. (\ref{eq:rhoT}) and (\ref{eq:rhoL}), all experimental data can be fitted simultaneously by the adjustable parameters $\theta_{SH}$, $\lambda$, $G_r$ and $G_i$. $\rho=1/\sigma$ follows from the measured resistances $R_0$ for each Pt thickness given in the caption of Fig. \ref{fig:Fig3}. The quality of the fit is demonstrated by Fig. \ref{fig:Fig3}(a)-(f) for $\theta_{SH}=0.08$, $\lambda=1.2$nm, $G_r=4.4\times10^{14}\Omega^{-1}$m$^{-2}$ and $G_i=2.8\times10^{13}\Omega^{-1}$m$^{-2}$. The measurements are very well described by the SMR theory (Eqs. (\ref{eq:rhoT}) and (\ref{eq:rhoL})), for all Pt-thicknesses and magnetic field strength and direction. However, due to the correlation between the fitting parameters, similarly good fitting results can be obtained by other combinations of $\theta_{SH}$, $\lambda$, $G_r$ and $G_i$, notwithstanding the good signal-to-noise-ratio of the experimental data. We therefore fixed the Hall angle at $\theta_{SH}=0.08$, which is within the range 0.06 to 0.11 obtained from the fitting and consistent with results published by several groups.\citep{ReviewSHE,LiuPRL2011,SHESaitohPRL,AzevedoPRB2011,VlietstraSMR} By fixing $\theta_{SH}$ the quality of the fits is not reduced, but the accuracy of the parameter estimations improves significantly. By Fig. \ref{fig:Fig4} it is observed that a strong correlation exists between both $G_r$ and $G_i$, and $\lambda$, whereas the ratio $G_r/G_i$ does not significantly change (see inset Fig. \ref{fig:Fig4}). A good fit cannot be obtained for $\lambda>1.4$nm. For $\lambda<0.8$nm the error bars become very large and for $\lambda<0.4$nm a good fit can no longer be obtained. Inspecting Fig. \ref{fig:Fig4} we favour $\lambda=1.1\pm0.3$nm, $G_r=(7\pm3)\times10^{14}\Omega^{-1}$m$^{-2}$ and $G_i=(5\pm3)\times10^{13}\Omega^{-1}$m$^{-2}$, where the higher values of $G_r$ and $G_i$ correspond to smaller $\lambda$. The ratio $G_r/G_i=16\pm4$ does not depend on $\lambda$.

In summary, by employing the SMR, including the contribution of the imaginary part of the spin-mixing conductance, it is possible to fully determine the magnetization direction of an insulating ferromagnetic layer, by purely electrical measurements. 
The experimental data are described well by the spin-diffusion model of the SMR, for all investigated Pt thicknesses and magnetic configurations. By fixing $\theta_{SH}=0.08$, we find the parameters $\lambda=1.1\pm0.3$nm, $G_r=(7\pm3)\times10^{14}\Omega^{-1}$m$^{-2}$, $G_i=(5\pm3)\times10^{13}\Omega^{-1}$m$^{-2}$ and $G_r/G_i=16\pm4$ for YIG/Pt bilayer structures.



We would like to acknowledge B. Wolfs, M. de Roosz and J. G. Holstein for technical assistance. This work is part of the research program of the Foundation for Fundamental Research on Matter (FOM), EU-ICT-7 "MACALO" and DFG Priority Programme 1538 "Spin-Caloric Transport" (BA 2954/1-1) and is supported by NanoNextNL, a micro and nanotechnology consortium of the Government of the Netherlands and 130 partners, by NanoLab NL and the Zernike Institute for Advanced Materials.

\providecommand{\noopsort}[1]{}\providecommand{\singleletter}[1]{#1}%
%



\end{document}